\documentstyle[12pt]{article}
\begin{document}
\textheight 22 cm
\baselineskip 18 pt plus2pt
\topmargin -30.0 mm
{\hfill  TECHNION--PH-95-5}

\vspace{6.0cm}
\centerline{\Large Dipole Perturbations of the Reissner-Nordstr\"{o}m}
\centerline{\Large Solution: The Polar Case}
\vspace{1.cm}
\centerline{Lior M. Burko, Department of Physics}
\centerline{Technion -- Israel
Institute of Technology, 32000 Haifa, Israel.}
\vspace{2.5cm}
\centerline{Abstract}
The formalism developed by Chandrasekhar
for the linear polar perturbations of the
Reissner-Nordstr\"{o}m solution is generalized to include the case of
dipole (l=1) perturbations. Then, the perturbed metric coefficients and
components of the Maxwell tensor are computed.

\vspace{2.5cm}
PACS numbers: 04.70.Bw \hspace{0.3cm} 04.25.Nx \hspace{0.3cm} 04.40.Nr
\newpage
\textheight 22 cm

\section{\sc Introduction}

The gravitational and electromagnetic perturbations of Schwarzschild and
Reissner-Nordstr\"{o}m black holes have been studied in detail
\cite{pert1,pert2,chandrasekhar,novikov - thorne}.
Even though
these two exact solutions are spherically symmetric, there is an important
difference in the character of their perturbations: In the Schwarzschild
solution the gravitational and electromagnetic (linear) perturbations are
uncoupled, while, in contradistinction, in the Reissner-Nordstr\"{o}m
solution, due to the background electric field, any electromagnetic
perturbation causes a gravitational perturbation, and {\it vice vers\^{a}}.
This coupling of the electromagnetic and the gravitational perturbations
complicates the study of the perturbations in Reissner-Nordstr\"{o}m black
holes considerably. Despite this complication,
it turns out that even in the Reissner-Nordstr\"{o}m black hole it
is possible to decouple the perturbations (of each multipole order and for
each parity) to two independent modes (each of
which is made of an electromagnetic component and a gravitational
component). This decoupling plays a crucial r\^{o}le in the study of
perturbations in Reissner-Nordstr\"{o}m black holes.

The decoupling of the perturbations of Reissner-Nordstr\"{o}m into
electromagnetic and metric perturbations was treated, for both polar and
axial modes, in Ref. \cite{pert2}  and summarized in Ref.
\cite{chandrasekhar}. (The treatment for the Schwarzschild black
hole is very similar, and is given in Refs. \cite{pert1} and
\cite{chandrasekhar}).
We shall see, however, that the formalism presented in Refs.
\cite{pert2,chandrasekhar} is not valid in the case of dipole $(l=1)$ modes.
For many applications, this
difficulty is not very crucial, as one may be primarily interested in the
dynamics of gravitational waves, for which there are no radiative modes
with $l<2$. However, it may be of interest to treat the propagation of
dipole electromagnetic waves, especially in the Reissner-Nordstr\"{o}m
spacetime, because of the coupling of the gravitational and the
electromagnetic fields. Thus, the late-time behavior of electromagnetic
perturbations produced during the collapse decays like the $(2l+2)$
inverse-power of external time \cite{price},
and is therefore dominated by the $l=1$ mode. In addition, the $l=1$
perturbations are especially important in the analysis of the
(electromagnetic) effects of the blue-sheet at the Cauchy horizon of
Reissner-Nordstr\"{o}m black holes \cite{burko and ori}. (Similar
electromagnetic effects are to be expected at the inner horizon of the Kerr
black hole, though we have not analyzed this case.)

In this Paper we modify the formalism given in
\cite{chandrasekhar} so that it can be applied for dipole polar modes.
Then, we generalize the formalism to include polar perturbations of any
$l$, including $l=1$. The treatment of axial perturbations is different and
we hope to treat them separately.

This Paper deals with perturbations of the Reissner-Nordstr\"{o}m black
hole. The perturbations of the Schwarzschild black hole are obtained from
our formalism as a special case. Throughout this Paper we shall use the
notation and convention of \cite{chandrasekhar} unless when explicitly
stated otherwise. As a rule, we shall not deviate from the notation of
\cite{chandrasekhar} except when necessary. When we do change the
notation, it will be by adding `bars' to the symbols of
\cite{chandrasekhar}. The `barred' objects will be defined such that they
are treated properly for dipole perturbations.

The outline of this Paper is as follows:
In Section 2 we shall describe the definitions
and notation. In Section 3 we give a full treatment for the general formalism
of polar perturbations of the Reissner-Nordstr\"{o}m solution. In Section 4
we shall decouple the fundamental equations for the perturbations for the
dipole case, and in Section 5 we shall generalize the treatment for all
polar modes. In Sections 6 and 7 we shall present the completion of the
solution, and in Section 8 we shall discuss the formalism and give some
concluding remarks.

\section{\sc Definitions and Notation}

Following Chandrasekhar \cite{chandrasekhar}, we
write the line-element of an unperturbed Reissner-Nordstr\"{o}m
black hole in the form
\begin{eqnarray}
\,ds^{2}&=&e^{2\nu}\left(\,dx^{0}\right)^{2}-e^{2\psi}\left(\,dx^{1}\right)
^{2}-e^{2\mu _{2}}\left(\,dx^{2}\right)^{2}-e^{2\mu _{3}}\left(
\,dx^{3}\right)^{2}\nonumber \\
&=&e^{2\nu}\left(\,dx^{0}\right)^{2}-e^{2\mu _{2}}\left(\,dx^{2}\right)^{2}
-r^{2}\,d\Omega^{2}, \end{eqnarray} where the co-ordinates are
\begin{eqnarray}
\left(x^{0}\;x^{1}\;x^{2}\;x^{3}\right)=\left(t\;\phi\;r\;\theta\right),
\end{eqnarray}
$\,d\Omega^{2}$ is the unit two-sphere line-element, and the metric
coefficients are $e^{2\nu}=e^{-2\mu_{2}}=(r^{2}-2Mr+Q_{*}^{2})/r^{2}
\equiv \Delta/r^{2}$,
$M,Q_{*}$ being the mass and electric charge, respectively, of the
Reissner-Nordstr\"{o}m black hole, and $r$ being the radial
Schwarzschild co-ordinate, defined such that circles of radius $r$ have
circumference $2\pi r$. The general form of the line-element (1) is
preserved under polar perturbations (sometimes called even-parity
perturbations); On the other hand, axial perturbations (called also odd-parity
perturbations), will lead in general to non-vanishing off-diagonal metric
coefficients\footnotemark \footnotetext{
Axial perturbations are characterized by the
non-vanishing of the metric functions $\omega, q_{2}, q_{3}$ (the
non-vanishing of these
metric-coefficients induce a dragging of the inertial-frame
and impart a rotation to the black hole), while polar
perturbations are those which alter the values of the metric functions
$\nu, \mu_{2}, \mu_{3}$ and $\psi$ (which are in general non-zero for the
unperturbed black hole).}. Therefore, the form of the metric of a
generally-perturbed Reissner-Nordstr\"{o}m black hole
will be much more complicated than the line-element (1).
It has been shown \cite{chandrasekhar},
that a metric of sufficient generality is of the form
\begin{eqnarray}
\,ds^{2}=e^{2\nu}\left(\,dx^{0}\right)^{2}&-&e^{2\psi}\left(\,dx^{1}-\omega
\,dx^{0}-q_{2}\,dx^{2}-q_{3}\,dx^{3}\right)^{2}\nonumber \\
&-&e^{2\mu_{2}}\left(\,dx^{2}\right)^{2}-e^{2\mu_{3}}
\left(\,dx^{3}\right)^{2}.
\end{eqnarray}
Since the unperturbed Reissner-Nordstr\"{o}m background is spherically
symmetric, we can consider only axisymmetric modes of perturbations without
any loss of generality\footnotemark
\footnotetext{This is because all non-axisymmetric modes can
be obtained from the axisymmetric modes, if the unperturbed spacetime is
spherically symmetric \cite{chandrasekhar}.}. The line-element (3) involves
seven  functions, namely, $\nu ,\psi ,\mu_{2} ,\mu_{3} ,\omega ,q_{2},$ and
$q_{3}$. Beacuse the Einstein equations involve only six independent
functions, not all seven functions can be determined arbitrarily, and
there is one constraint on the metric coefficient. It has been
shown \cite{chandrasekhar}, that this constraint is
$$(\omega_{,2}-q_{2,0})_{,3}-(\omega_{,3}-q_{3,0})_{,2}+(q_{2,3}-q_{3,2})
_{,0}.$$

\section{\sc The General Formalism}

For completeness, we shall first present the linearized field-equations and
the decoupling of the $r,\theta$ varibles as given in \cite{chandrasekhar}:
The formalism for the treatment of the perturbations is made
of the linearization of the
coupled Einstein-Maxwell equations about the Reissner-Nordstr\"{o}m
solution. In particular, linearization of the Ricci, Einstein and Maxwell
tensors leads to the following equations \cite{chandrasekhar}:
\begin{eqnarray}
\left(\,\delta\psi +\,\delta\mu_{3}\right)_{,r}+\left(\frac{1}{r}-
\nu_{,r}\right)\left(\,\delta\psi +\,\delta\mu_{3}\right)-
\frac{2}{r}\,\delta\mu_{2}=-\,\delta R_{(0)(2)}=0,
\end{eqnarray}
\begin{eqnarray}
\left[\left(\,\delta\psi +\,\delta\mu_{2}\right)_{,\theta}+
\left(\,\delta\psi -\,\delta\mu_{3}\right)\cot\theta\right]_{,0}=
-e^{\nu+\mu_{3}}\,\delta R_{(0)(3)}=-2Q_{*}\frac{e^{\nu}}{r}
F_{(2)(3)},
\end{eqnarray}
\begin{eqnarray}
\left(\,\delta\psi +\,\delta\nu\right)_{r,\theta}&+&
\left(\,\delta\psi -\,\delta\mu_{3}\right)_{,r}\cot\theta -
\left(\frac{1}{r}-\nu_{,r}\right)\,\delta\nu_{,\theta}-
\left(\frac{1}{r}+\nu_{,r}\right)\,\delta\mu_{2,\theta}\nonumber\\
&=&-e^{\mu_{2}+\mu_{3}}\,\delta R_{(2)(3)}=-2Q_{*}\frac{e^{-\nu}}{r}
F_{(0)(3)},
\end{eqnarray}
\begin{eqnarray}
e^{2\nu}\left[\frac{2}{r}\,\delta\nu_{,r}\right.&+&
\left.\left(\frac{1}{r}+\nu_{,r}\right)
\left(\,\delta\psi +\,\delta\mu_{3}\right)_{,r}-2\left(
\frac{1}{r^{2}}+2\frac{\nu_{,r}}{r}\right)\,\delta\mu_{2}\right]\nonumber\\
&+&\frac{1}{r^{2}}\left[(\,\delta\psi +\,\delta\nu )
_{,\theta,\theta}+(2\,\delta\psi+\,\delta\nu-\,\delta\mu_{3})
_{,\theta}\cot\theta+2\,\delta\mu_{3}\right]\nonumber\\
&-&e^{-2\nu}(\,\delta\psi +\,\delta\mu_{3})_{,0,0}\nonumber\\
&=&\,\delta G_{(2)(2)}=\,\delta R_{(2)(2)}=2\frac{Q_{*}}{r^{2}}\,\delta
F_{(0)(2)},
\end{eqnarray}
\begin{eqnarray}
&e^{2\nu}&\left[\,\delta\psi_{,r,r}+
2\left(\frac{1}{2}+\nu_{,r}\right)+
\frac{1}{r}\left(\,\delta\psi+\,\delta\nu+\,\delta\mu_{3}-\,\delta\mu_{2}
\right)_{,r}\right.\nonumber\\
&-&\left. 2\left(\frac{1}{2}+2\nu_{,r}\right)\frac{1}{r}\,\delta\mu_{2}
\right]+\frac{1}{r^{2}}[\,\delta\psi_{,\theta ,\theta}+
\,\delta\psi_{,\theta}\cot\theta+
\left(\,\delta\psi+\,\delta\nu\right.\nonumber\\
&-&\left.\,\delta\mu_{3}+\,\delta\mu_{2}\right)
_{,\theta}\cot\theta+2\,\delta\mu_{3}]-
e^{-2\nu}\,\delta\psi_{,0,0}\nonumber\\&=&-\,\delta R_{(1)(1)}=
2\frac{Q_{*}}{r^{2}}\,\delta F_{(0)(2)},
\end{eqnarray}
\begin{eqnarray}
re^{-\nu}F_{(0)(3),0}=\left[re^{\nu}F_{(2)(3)}\right]_{,r},
\end{eqnarray}
\begin{eqnarray}
\,\delta F_{(0)(2),0}-\frac{Q_{*}}{r^{2}}\left(\,\delta\psi
+\,\delta\mu_{3}\right)_{,0}+\frac{e^{\nu}}{r\sin\theta}
\left[F_{(2)(3)}\sin\theta\right]_{,\theta}=0,
\end{eqnarray}
\begin{eqnarray}
\left[\,\delta F_{(0)(2)}-\frac{Q_{*}}{r^{2}}
\left(\,\delta\nu+\,\delta\mu_{2}\right)\right]_{,\theta}
+\left[re^{\nu}F_{(3)(0)}\right]_{,r}+re^{-\nu}F_{(2)(3),0}=0,
\end{eqnarray}
where $A_{(\alpha)(\beta)}$ is the $\alpha\beta$ tetrad component of the
tensor $A$, and $F,G,$ and $R$ are the Maxwell, Einstein, and Ricci
tensors, respectively. The variables $r$ and $\theta$ in Eqs. (4)--(11) can
be separated by the Friedman substitutions \cite{friedman}
\begin{eqnarray}
\,\delta\nu&=&N(r)P_{l}(\cos\theta),\\
\,\delta\mu_{2}&=&L(r)P_{l}(\cos\theta),\\
\,\delta\mu_{3}&=&\left[T(r)P_{l}(\cos\theta)+V(r)P_{l,\theta,\theta}
(\cos\theta)\right],\\
\,\delta\psi&=&\left[T(r)P_{l}(\cos\theta)+V(r)P_{l,\theta}(\cos\theta)
\cot\theta\right],\\
\,\delta F_{(0)(2)}&=&\frac{r^{2}e^{2\nu}}{2Q_{*}}B_{(0)(2)}(r)P_{l}
(\cos\theta),\\
F_{(0)(3)}&=&\frac{r^{2}e^{\nu}}{2Q_{*}}B_{(0)(3)}(r)
P_{l,\theta}(\cos\theta),
\end{eqnarray}
and
\begin{eqnarray}
F_{(2)(3)}=-i\sigma\frac{r^{2}e^{-\nu}}{2Q_{*}}B_{(2)(3)}(r)
P_{l,\theta}(\cos\theta).
\end{eqnarray}
$P_{l}(\cos\theta)$ are the Legendre functions of order $l$. We
assume that the perturbations can be analyzed into their normal modes with
a time dependence $e^{i\sigma t}$. This Fourier decomposition of the
perturbations can be done without any loss of generality due to the
linearized theory we assume\footnotemark \footnotetext{
We note, that these components are frequency-dependent. To obtain
components independent of the frequency one should Fourier-transform from the
frequency-plane to the temporal-plane.}. Using these substitutions, we obtain
the following equations for the radial functions defined by Eqs.
(12)--(18):
\begin{eqnarray}
\left[\frac{d}{\,dr}+\left(\frac{1}{r}-\nu_{,r}\right)\right]
\left[2T-l(l+1)V\right]-\frac{2}{r}L=0,\\
(T-V+L)=B_{(2)(3)},\\
(T-V+N)_{,r}-\left(\frac{1}{r}-\nu_{,r}\right)N-
\left(\frac{1}{r}+\nu_{,r}\right)L=B_{(0)(3)},
\end{eqnarray}
\begin{eqnarray}
\frac{2}{r}N_{,r}&+&\left(\frac{1}{r}+\nu_{,r}\right)
\left[2T-l(l+1)V\right]-\frac{2}{r}\left(\frac{1}{r}+2\nu_{,r}\right)L
\nonumber\\
&-&\frac{l(l+1)}{r^{2}}e^{-2\nu}N-\frac{(l-1)(l+2)}{r^{2}}
e^{-2\nu}T+\sigma^{2}e^{-4\nu}\left[2T-l(l+1)V\right]\nonumber\\
&=&B_{(0)(2)},
\end{eqnarray}
\begin{eqnarray}
B_{(0)(3)}=\frac{1}{r^{2}}\left[r^{2}B_{(2)(3)}\right]_{,r}
=B_{(2)(3),r}+\frac{2}{r}B_{(2)(3)},\\
r^{4}e^{2\nu}B_{(0)(2)}=2Q_{*}^{2}\left[2T-l(l+1)V\right]-l(l+1)r^{2}
B_{(2)(3)},\\
\left[r^{2}e^{2\nu}B_{(0)(3)}\right]_{,r}+r^{2}e^{2\nu}B_{(0)(2)}+
\sigma^{2}r^{2}e^{-2\nu}B_{(2)(3)}=2Q_{*}^{2}\frac{N+L}{r^{2}}.
\end{eqnarray}
Note, that in Eq. (22) we changed the formalism of \cite{chandrasekhar}.
We shall now see the reasons for this change in the formalism, which makes
the extension of the formalism to $l=1$ necessary.
In Ref. \cite{chandrasekhar},
a new radial function $X$ is defined by
\begin{eqnarray}
X=nV=\frac{1}{2}(l-1)(l+2)V.\nonumber
\end{eqnarray}
For dipole radiation $n$, and consequently $X$, vanish. Hence, it is clear
that the variable $X$ -- because it vanishes identically for dipole
perturbations -- cannot carry any information on the original variable $V$,
which is to be calculated. As it is clear that the perturbative terms do
not vanish identically (it is well known that there is in general a dipole
electromagnetic mode, also in Minkowski spacetime), the formalism of
\cite{chandrasekhar} needs to be generalized to be valid for the treatment
of dipole radiation too. Furthermore, in Ref. \cite{chandrasekhar}
physically-meaningful
variables are devided by $n$ or by $\mu$, where $\mu^{2}\equiv 2n$. This is
clearly inappropriate\footnotemark\footnotetext{
See, e.g., Eqs. (180)--(181) of Chapter 5 of \cite{chandrasekhar}.}
for dipole radiation due to the unity value of $l$
and consequently the identically-vanishing values of $n$ and $\mu$.

We re-write Eq. (20) as
\begin{eqnarray}
2T-l(l+1)V=-2\left[L+\frac{1}{2}(l-1)(l+2)V-B_{(2)(3)}\right],
\end{eqnarray}
which, after substitution in Eq. (19) yields
\begin{eqnarray}
\left[L+\frac{1}{2}(l-1)(l+2)V-B_{(2)(3)}\right]_{,r}=&-&\left(\frac{1}{r}-
\nu_{,r}\right)\left[L+\frac{1}{2}(l-1)(l+2)V\right.\nonumber\\
&-&\left.B_{(2)(3)}\right]-\frac{1}{r}L.
\end{eqnarray}
Combining Eqs. (20),(21), and (23) we obtain
\begin{eqnarray}
\left(N-L\right)_{,r}=\left(\frac{1}{r}-\nu_{,r}\right)N+
\left(\frac{1}{r}+\nu_{,r}\right)L+\frac{2}{r}B_{(2)(3)}
\end{eqnarray}
{}From Eqs. (22),(27), and (28) we find the following equations for the
radial functions $L$,$N$, and $V$: (Note, that Eq. (31) is an equation for
the variable $V$, which replaces the equation for $X$ given in
\cite{chandrasekhar}. For dipole radiation $X$ vanishes identically, and
therefore is not to be treated.)
\begin{eqnarray}
N_{,r}&=&aN+bL+c\left[\frac{1}{2}(l-1)(l+2)V-B_{(2)(3)}\right],\\
L_{,r}&=&\left(a-\frac{1}{r}+\nu_{,r}\right)N+
\left(b-\frac{1}{r}-\nu_{,r}\right)L\nonumber\\&+&c\left[
\frac{1}{2}(l-1)(l+2)V-B_{(2)(3)}\right]-\frac{2}{r}B_{(2)(3)},\\
\frac{1}{2}(l-1)(l+2)V_{,r}=&-&\left(a-\frac{1}{r}+\nu_{,r}\right)N-
\left(b+\frac{1}{r}-2\nu_{,r}\right)L\nonumber\\
&-&\left(c+\frac{1}{r}
-\nu_{,r}\right)\left[\frac{1}{2}(l-1)(l+2)V-B_{(2)(3)}\right]
\nonumber\\&+&B_{(0)(3)},
\end{eqnarray}
where
\begin{eqnarray}
a&=&\frac{1+(l-1)(l+2)/2}{r}e^{-2\nu},\\
b&=&-\frac{1}{r}-\left[\frac{(l-1)(l+2)}{2r}-\frac{M}{r^{2}}
\right]e^{-2\nu}\nonumber\\&+&\left[\frac{M^{2}}{r^{3}}+\sigma^{2}r-
\frac{Q_{*}^{2}}{r^{3}}\left(1+2e^{2\nu}\right)\right]e^{-4\nu},\\
c&=&-\frac{1}{r}+\frac{1}{r}e^{-2\nu}+
\left[\frac{M^{2}}{r^{3}}+\sigma^{2}r-
\frac{Q_{*}^{2}}{r^{3}}\left(1+2e^{2\nu}\right)\right]e^{-4\nu}.
\end{eqnarray}
It is important to notice, that for dipole radiation Eq. (31) becomes an
algebraic equation rather than a differential equation. (We shall see this
in detail when we explicitly discuss the dipole mode.) Eqs.
(23),(25),(29),(30), and (31) can be reduced to a pair of second-order
equations (and thus allow for a special solution \cite{chandrasekhar}).
We now define the following functions: (Notice the difference
between these functions and the functions defined in Ref.
\cite{chandrasekhar}.)
\begin{eqnarray}
\bar{H}_{2}^{(+)}&=&rV-\frac{r^{2}}{\varpi}\left[L+\frac{1}{2}(l-1)(l+2)V
-B_{(2)(3)}\right],\\
\bar{H}_{1}^{(+)}&=&-\frac{1}{Q_{*}}\left\{r^{2}B_{(2)(3)}+
2Q_{*}^{2}\frac{r}{\varpi}\left[L+\frac{(l-1)(l+2)}{2}V-B_{(2)(3)}\right]
\right\},
\end{eqnarray}
where $\varpi=(l-1)(l+2)r/2+3M-2Q_{*}^{2}/r.$ The newly-defined functions
satisfy the following coupled equations:
\begin{eqnarray}
\Lambda^{2}\bar{H}_{2}^{(+)}&=&\frac{\Delta}{r^{5}}\left\{\bar{U}
\bar{H}_{2}^{(+)}+\bar{W}\left[-3M\bar{H}_{2}^{(+)}+2Q_{*}\bar{H}_{1}^{(+)}
\right]\right\},\\
\Lambda^{2}\bar{H}_{1}^{(+)}&=&\frac{\Delta}{r^{5}}\left\{\bar{U}
\bar{H}_{1}^{(+)}+\bar{W}\left[2Q_{*}(l-1)(l+2)\bar{H}_{2}^{(+)}
+3M\bar{H}_{1}^{(+)}\right]\right\},
\end{eqnarray}
where
\begin{eqnarray}
\bar{U}&=&\left[(l-1)(l+2)r+3M\right]\bar{W}+\left[\varpi-(l-1)(l+2)r
-M\right]\nonumber\\&-&\frac{(l-1)(l+2)\Delta}{\varpi},\\
\bar{W}&=&\frac{\Delta}{r\varpi^{2}}\left[(l-1)(l+2)r+3M\right]
+\frac{(l-1)(l+2)r+M}{\varpi},
\end{eqnarray}
and $\Lambda^{2}\equiv d^{2}/\,dr_{*}^{2}+\sigma^{2}$,~ $r_{*}$ being the
Regge-Wheeler `tortoise' co-ordinate defined by
$(\Delta /r^{2})d/\,dr=d/\,dr_{*}$.

\section {\sc Decoupling of the Equations -- Dipole Case}
The decoupling of the equations for the radial functions
$\bar{H}_{1}^{(+)},\bar{H}_{2}^{(+)}$ is easier when one first decouples
them for the special case $l=1$, and then uses this case
for the determination of parameters for the decoupling of the general
equations. In the next section we shall decouple
the equations for any $l$. for $l=1$, Eqs. (37) and (38) assume the form
\begin{eqnarray}
\Lambda^{2}\bar{H}_{1}^{(+)}&=&\frac{\Delta}{r^{5}}\left(\bar{U}+3M\bar{W}
\right)\bar{H}_{1}^{(+)}, \\
\Lambda^{2}\bar{H}_{2}^{(+)}&=&\frac{\Delta}{r^{5}}\left[\left(\varpi -M
\right)\bar{H}_{2}^{(+)}+2Q_{*}\bar{W}\bar{H}_{1}^{(+)}\right].
\end{eqnarray}
It is important to notice, that Eq. (41) is already decoupled. We shall
find it convenient to
define new radial functions $\bar{Z}_{1}^{(+)},\bar{Z}_{2}^{(+)}$ by
\begin{eqnarray}
\bar{H}_{1}^{(+)}&=&\alpha\bar{Z}_{1}^{(+)}+\beta\bar{Z}_{2}^{(+)}, \\
\bar{H}_{2}^{(+)}&=&\gamma\bar{Z}_{1}^{(+)}+\delta\bar{Z}_{2}^{(+)}.
\end{eqnarray}
Because Eq. (41) is decoupled, we find that for $l=1$, ~ $\beta =0$.
Substituting Eqs. (43) and (44) in Eqs. (41) and (42), we find that
\begin{eqnarray}
\alpha\Lambda^{2}\bar{Z}_{1}^{(+)}&=&\alpha\frac{\Delta}{r^{5}}
\left(\bar{U}+3M\bar{W}\right)\bar{Z}_{1}^{(+)}, \\
\gamma\Lambda^{2}\bar{Z}_{1}^{(+)}+\delta\Lambda^{2}\bar{Z}_{2}^{(+)}
&=&\frac{\Delta}{r^{5}}\left[\gamma (\varpi -M)+2\alpha Q_{*}\bar{W}\right]
\bar{Z}_{1}^{(+)}\nonumber \\
&+&\frac{\Delta}{r^{5}}\delta (\varpi -M)\bar{Z}_{2}^{(+)}.
\end{eqnarray}
Multiplying Eq. (45) by $\gamma$, Eq. (46) by $\alpha$, and substracting
the resultant equations, we find that
\begin{eqnarray}
\alpha\delta\Lambda^{2}\bar{Z}_{2}^{(+)}&=&\alpha\frac{\Delta}{r^{5}}
\left[\gamma (\varpi -M)+2\alpha Q_{*}\bar{W}-\gamma (\bar{U}+3M\bar{W})
\right]\bar{Z}_{1}^{(+)} \nonumber \\
&+&\alpha\delta\frac{\Delta}{r^{5}}(\varpi -M)\bar{Z}_{2}^{(+)}.
\end{eqnarray}
In order that Eq. (47) indeed be decoupled,
the decoupling parameters $\alpha$ and $\gamma$ must be such that
\begin{eqnarray}
\gamma (\varpi -M)+2\alpha Q_{*}\bar{W}-\gamma (\bar{U}+3M\bar{W})=0.
\end{eqnarray}
We thus obtain that
\begin{eqnarray}
\alpha=\frac{3M\bar{W}+\bar{U}+M-\varpi}{2Q_{*}\bar{W}}\gamma,
\end{eqnarray}
or, substituting Eq. (39) for $\bar{U}$,
\begin{eqnarray}
\alpha =3\frac{M}{Q_{*}}\gamma .
\end{eqnarray}
We still have the freedom to fix one of the parameters $\alpha$ or
$\gamma$. Choosing $\alpha =1/(6M)$ [and, consequently, $\gamma =
Q_{*}/(18M^{2})$], we obtain for the decoupled equations (in the $l=1$
case):
\begin{eqnarray}
\Lambda^{2}\bar{Z}_{1}^{(+)}&=&\frac{\Delta}{r^{5}}\left(2M-2\frac{Q_{*}^{2}}
{r}+6M\bar{W}\right)\bar{Z}_{1}^{(+)}, \\
\Lambda^{2}\bar{Z}_{2}^{(+)}&=&\frac{\Delta}{r^{5}}\left(2M-2\frac{Q_{*}^{2}}
{r}\right)\bar{Z}_{2}^{(+)}.
\end{eqnarray}
We notice, that $\delta$ remains free to be fixed arbitrarily.
\section {\sc Decoupling of the Equations -- General Case}
In this section, we shall decouple Eqs. (37) and (38) for any $l$. We again
use Eqs. (43) and (44), but in this case, of course, $\beta$ will in
general not vanish identically.  We thus find that
\begin{eqnarray}
\alpha\Lambda^{2}\bar{Z}_{1}^{(+)}&+&\beta\Lambda^{2}\bar{Z}_{2}^{(+)}=
\frac{\Delta}{r^{5}}\left[\alpha\bar{U}+2\gamma Q_{*}(l-1)(l+2)\bar{W}
+3\alpha M\bar{W}\right]\bar{Z}_{1}^{(+)}\nonumber \\
&+&\frac{\Delta}{r^{5}}\left[\beta\bar{U}+2\delta Q_{*}(l-1)(l+2)\bar{W}
+3\beta M\bar{W}\right]\bar{Z}_{2}^{(+)},\\
\gamma\Lambda^{2}\bar{Z}_{1}^{(+)}&+&\delta\Lambda^{2}\bar{Z}_{2}^{(+)}=
\frac{\Delta}{r^{5}}(\gamma\bar{U}-3\gamma M\bar{W}+2\alpha Q_{*}\bar{W})
\bar{Z}_{1}^{(+)}\nonumber \\
&+&\frac{\Delta}{r^{5}}(\delta\bar{U}-3\delta M\bar{W}+2\beta Q_{*}
\bar{W})\bar{Z}_{2}^{(+)}.
\end{eqnarray}
We now multiply Eq. (53) by $\gamma$ and Eq. (54) by $\alpha$. Substracting
the equations we find that
\begin{eqnarray}
(\beta\gamma &-&\alpha\delta )\Lambda^{2}\bar{Z}_{2}^{(+)}=\frac{\Delta}
{r^{5}}\left[2\gamma^{2}Q_{*}(l-1)(l+2)\bar{W}+6\alpha\gamma M\bar{W}
\right.\nonumber\\
&-&\left. 2\alpha^{2}Q_{*}\bar{W}\right]\bar{Z}_{1}^{(+)}
+\frac{\Delta}{r^{5}}\left[\beta\gamma\bar{U}+2\gamma\delta Q_{*}
(l-1)(l+2)\bar{W}+3\beta\gamma M\bar{W}\right.\nonumber\\
&-&\left.\alpha\delta\bar{U}+3\alpha\delta M
\bar{W}-2\alpha\beta Q_{*}\bar{W}\right]\bar{Z}_{2}^{(+)}.
\end{eqnarray}
We now require that
\begin{eqnarray}
2\gamma^{2}Q_{*}(l-1)(l+2)+6\alpha\gamma M-2\alpha^{2}Q_{*}=0.
\end{eqnarray}
The solution of this constraint is
\begin{eqnarray}
\alpha=\frac{\gamma\left[3M\pm \sqrt{9M^{2}+4Q_{*}^{2}(l-1)(l+2)}\right]}
{2Q_{*}}.
\end{eqnarray}
To obtain the result of the previous section
for the $l=1$ mode, we choose the positive root. We now define
\begin{eqnarray}
q_{1}=3M+\sqrt{9M^{2}+4Q_{*}^{2}(l-1)(l+2)},
\end{eqnarray}
and find that
\begin{eqnarray}
\alpha =\frac{q_{1}}{2Q_{*}}\gamma.
\end{eqnarray}
For the $l=1$ case we find that $q_{1}=6M$, and we thus indeed recover our
previous result for the $l=1$ case [Eq. (50)].
To obtain a corresponding connection between $\beta$ and $\delta$ we
multiply Eq.
(53) by $\delta$ and Eq. (54) by $\beta$. Substracting the equations we
find that
\begin{eqnarray}
(\beta\gamma&-&\alpha\delta)\Lambda^{2}\bar{Z}_{1}^{(+)}=
\frac{\Delta}{r^{5}}\left[\beta\gamma\bar{U}-3\beta\gamma M\bar{W}+
2\alpha\beta Q_{*}\bar{W}-\alpha\delta\bar{U}\right.\nonumber \\
&-&2\gamma\delta (l-1)(l+2)Q_{*}\bar{W}
-\left.3\alpha\delta M\bar{W}\right]\bar{Z}_{1}^{(+)}+
\frac{\Delta}{r^{5}}\left[-6\beta\delta
M\bar{W}\right.\nonumber \\
&+&\left.2\beta^{2}Q_{*}\bar{W}-2\delta^{2}(l-1)(l+2)
Q_{*}\bar{W}\right]\bar{Z}_{2}^{(+)}.
\end{eqnarray}
To allow for the decoupling we require that
\begin{eqnarray}
2\beta^{2}Q_{*}-2\delta^{2}(l-1)(l+2)Q_{*}-6\beta\delta M=0.
\end{eqnarray}
The solution of Eq. (61) is:
\begin{eqnarray}
\beta=\frac{\delta\left[3M\pm\sqrt{9M^{2}+4(l-1)(l+2)Q_{*}^{2}}
\right]}{2Q_{*}}.
\end{eqnarray}
Because for $l=1$ we have $\beta =0$, we choose the negative root, and
define
\begin{eqnarray}
q_{2}=3M-\sqrt{9M^{2}+4Q_{*}^{2}(l-1)(l+2)}.
\end{eqnarray}
We thus find that
\begin{eqnarray}
\beta=\frac{q_{2}}{2Q_{*}}\delta .
\end{eqnarray}
We now fix $\delta =1/q_{1}$, and consequently $\beta =q_{2}/
(2q_{1}Q_{*})$. Thus, we found the four parameters of Eqs. (43) and (44), and
completed the decoupling of the equations.

\section{\sc The Decoupled Equations}
In the previous section we found that Eqs. (43) and (44) can be explicitly
written as
\begin{eqnarray}
\bar{H}_{1}^{(+)}&=&\frac{1}{q_{1}}\bar{Z}_{1}^{(+)}+\frac{q_{2}}
{2q_{1}Q_{*}}\bar{Z}_{2}^{(+)},\\
\bar{H}_{2}^{(+)}&=&\frac{Q_{*}}{3Mq_{1}}\bar{Z}_{1}^{(+)}+
\frac{1}{q_{1}}\bar{Z}_{2}^{(+)}.
\end{eqnarray}
Substituting Eqs. (65) and (66) in Eqs. (37) and (38) we find that the
differential equations
satisfied by $\bar{Z}_{1}^{(+)}$, ~$\bar{Z}_{2}^{(+)}$ are
\begin{eqnarray}
\Lambda^{2}\bar{Z}_{1}^{(+)}=\frac{\Delta}{r^{5}}\left\{\bar{U}\right.&+&
\left.\frac{9M^{2}\bar{W}}{\sqrt{9M^{2}+4Q_{*}^{2}(l-1)(l+2)}}\right.\nonumber\\
&+&\left.
\frac{[q_{1}q_{2}-4Q_{*}^{2}(l-1)(l+2)]\bar{W}}
{q_{2}-q_{1}}\right\}
\bar{Z}_{1}^{(+)},
\end{eqnarray}
\begin{eqnarray}
\Lambda^{2}\bar{Z}_{2}^{(+)}=\frac{\Delta}{r^{5}}\left\{\bar{U}\right.&-&
\left.\frac{9M^{2}\bar{W}}{\sqrt{9M^{2}+4Q_{*}^{2}(l-1)(l+2)}}\right.\nonumber\\
&-&\left.
\frac{[q_{1}q_{2}-4Q_{*}^{2}(l-1)(l+2)]\bar{W}}{q_{2}-q_{1}}\right\}
\bar{Z}_{2}^{(+)}.
\end{eqnarray}
Eqs. (67) and (68) can be re-written as
\begin{eqnarray}
\Lambda^{2}\bar{Z}_{i}^{(+)}=\bar{V}_{i}^{(+)}\bar{Z}_{i}^{(+)},
\end{eqnarray}
where $i=1,2$ and
\begin{eqnarray}
\bar{V}_{1,2}^{(+)}=\frac{\Delta}{r^{5}}\left[\bar{U}\pm\frac{1}{2}
\left(q_{1}-q_{2}\right)\bar{W}\right].
\end{eqnarray}

\section{\sc The Completion of the Solution}

As five differential equations of the first order are reduced to a pair
of second-order equations, it is clear that there is a special solution.
This special solution is \cite{chandrasekhar}:
\begin{eqnarray}
N^{(0)}&=&r^{-2}e^{\nu}\left[M-\frac{r}{\Delta}\left(M^{2}-Q_{*}^{2}+
\sigma^{2}r^{4}\right)-2\frac{Q_{*}^{2}}{r}\right]\\
L^{(0)}&=&r^{-3}e^{\nu}\left(3Mr-4Q_{*}^{2}\right)\\
V^{(0)}&=&e^{\nu}r^{-1}\\
B_{(2)(3)}^{(0)}&=&-2Q_{*}^{2}r^{-3}e^{\nu}\\
B_{(0)(3)}^{(0)}&=&2Q_{*}^{2}r^{-6}e^{-\nu}\left(2Q_{*}^{2}+r^{2}-3Mr\right).
\end{eqnarray}
As in Ref. \cite{chandrasekhar}, the completion of the solution is given by:
\begin{eqnarray}
N&=&N^{(0)}\Phi+(l-1)(l+2)\frac{e^{2\nu}}{\varpi}\bar{H}_{2}^{(+)}
-\frac{e^{2\nu}}{\varpi}\left[\frac{1}{2}(l-1)(l+2)r\bar{H}_{2}^{(+)}
\right.\nonumber\\
&+&\left.Q_{*}\bar{H}_{1}^{(+)}\right]_{,r}+
\frac{1}{r\varpi^{2}}\left\{e^{2\nu}\left[\varpi -(l-1)(l+2)r-3M\right]
\right.\nonumber\\
&-&\left.\left[\frac{1}{2}(l-1)(l+2)+1\right]\varpi
\right\}\left[\frac{1}{2}(l-1)(l+2)r\bar{H}_{2}^{(+)}+
Q_{*}\bar{H}_{1}^{(+)}\right]\\
L&=&L^{(0)}\Phi-\frac{1}{r^{2}}\left[\frac{1}{2}(l-1)(l+2)
r\bar{H}_{2}^{(+)}+Q_{*}\bar{H}_{1}^{(+)}\right]\\
V&=&V^{(0)}\Phi+\frac{1}{r}\bar{H}_{2}^{(+)}\\
B_{(2)(3)}&=&B_{(2)(3)}^{(0)}\Phi-\frac{Q_{*}}{r^{2}}\bar{H}_{1}^{(+)}\\
B_{(0)(3)}&=&B_{(0)(3)}^{(0)}\Phi-\frac{Q_{*}}{r^{2}}\bar{H}_{1,r}^{(+)}
\nonumber\\&-&2\frac{Q_{*}^{2}}{r^{2}\varpi}
\left[\frac{1}{2}(l-1)(l+2)r\bar{H}_{2}^{(+)}+Q_{*}\bar{H}_{1}^{(+)}
\right]\\
T&=&B_{(2)(3)}+V-L\\
B_{(0)(2)}&=&r^{-4}e^{-2\nu}\left\{ 2Q_{*}^{2}\left[2T-l(l+1)V\right]
-l(l+1)r^{2}B_{(2)(3)}\right\},
\end{eqnarray}
where
\begin{eqnarray}
\Phi=\int\left[\frac{1}{2}(l-1)(l+2)r\bar{H}_{2}^{(+)}+Q_{*}\bar{H}_{1}
^{(+)}\right]\frac{e^{-\nu}}{\varpi r}\,dr.
\end{eqnarray}

\section{\sc Discussion}

The formalism presented here
is adequate for the treatment of polar modes of any $l$,
including $l=1$. Now, we
shall see in detail the perturbation formalism for dipole polar
perturbations. We can simply substitute a value of unity for $l$,
and obtain the
equations for the dipole mode. We observe, that Eq. (31) becomes an
algebraic equation rather than a differential equation. This results from
the non-radiative character of the dipole gravitational mode. Hence,
dynamics is obtained from just one differential equation (of the second
order) and not by a pair of second-order differential equations. The
expression for Eq. (31) in the case of dipole perturbations then reads
\begin{eqnarray}
&-&\left(a-\frac{1}{r}+\nu_{,r}\right)N-\left(b+\frac{1}{r}-2\nu_{,r}\right)L
\nonumber\\
&+&\left(c+\frac{1}{r}-\nu_{,r}\right)B_{(2)(3)}+B_{(0)(3)}=0,
\end{eqnarray}
with $a,b$ and $c$ defined by Eqs. (32)--(34).
Now, we re-write Eqs. (35) and (36) as
\begin{eqnarray}
\bar{H}_{2}^{(+)}(l=1)=rV-\frac{r^{2}}{\varpi}(L-B_{(2)(3)}),
\end{eqnarray}
and
\begin{eqnarray}
\bar{H}_{1}^{(+)}(l=1)=-\frac{1}{Q_{*}}\left\{r^{2}B_{(2)(3)}+2Q_{*}^{2}
\frac{r}{\varpi}\left[L-B_{(2)(3)}\right]\right\},
\end{eqnarray}
where $\varpi (l=1)=3M-2Q_{*}^{2}/r$.
With these definitions, the differential
equation for $\bar{H}_{1}^{(+)}(l=1)$
is already decoupled from the equation for
$\bar{H}_{2}^{(+)}(l=1)$, and it reads
\begin{eqnarray}
\Lambda^{2}\bar{H}_{1}^{(+)}(l=1)=\frac{\Delta}{r^{5}}\left(\tilde{U}+3M
\tilde{W}\right)\bar{H}_{1}^{(+)}(l=1),
\end{eqnarray}
where $\tilde{U}=3M\tilde{W}+(\varpi-M)$ and
$\tilde{W}=3M\Delta/(r\varpi^{2})+M/\varpi$.
It turns out, that all the physically meaningful quantities are fully
determined by $\bar{H}_{1}^{(+)}$.
The completion of the solution is now given by the following relations:
\begin{eqnarray}
N&=&N^{(0)}\Phi-\frac{e^{2\nu}}
{\varpi}Q_{*}\bar{H}_{1,r}^{(+)}+
\frac{Q_{*}}{r\varpi^{2}}\left[e^{2\nu}\left(\varpi -3M\right)-\varpi
\right] \bar{H}_{1}^{(+)}\\
L&=&L^{(0)}\Phi-\frac{Q_{*}}{r^{2}}\bar{H}_{1}^{(+)}\\
B_{(2)(3)}&=&B_{(2)(3)}^{(0)}\Phi-\frac{Q_{*}}{r^{2}}\bar{H}_{1}^{(+)}\\
B_{(0)(3)}&=&B_{(0)(3)}^{(0)}\Phi-\frac{Q_{*}}{r^{2}}\bar{H}_{1,r}^{(+)}
-2\frac{Q_{*}^{3}}{r^{2}\varpi}\bar{H}_{1}^{(+)}\\
B_{(0)(2)}&=&r^{-4}e^{-2\nu}\left\{ 4Q_{*}^{2}[B_{(2)(3)}-L]
-2r^{2}B_{(2)(3)}\right\},
\end{eqnarray}
where we use Eqs. (47)--(51)
for the definitions of the special functions used in
the above equations. The function $\Phi (l=1)$ is
\begin{eqnarray}
\Phi (l=1)=Q_{*}\int\bar{H}_{1}^{(+)}\frac{e^{-\nu}}{\varpi r}\,dr .
\end{eqnarray}
Using these radial functions, the metric perturbations [through Eqs.
(12)--(15)]
are given by
\begin{eqnarray}
\,\delta\nu (l=1)&=&N(r)\cos\theta \\
\,\delta\mu_{2} (l=1)&=&L(r)\cos\theta \\
\,\delta\mu_{3} (l=1)&= &\delta\psi \nonumber \\
&=&\left[B_{(2)(3)}-L\right]\cos\theta ,
\end{eqnarray}
and the perturbations of the tetrad components of the Maxwell
tensor [Eqs. (16)--(18)] are
\begin{eqnarray}
\,\delta F_{(0)(2)} (l=1)&=&\frac{r^{2}e^{2\nu}}{2Q_{*}}B_{(0)(2)}
(r)\cos\theta\\
F_{(0)(3)} (l=1)&=&\frac{re^{\nu}}{2Q_{*}}B_{(0)(3)}(r)\sin\theta\\
F_{(2)(3)} (l=1)&=&i\sigma\frac{re^{-\nu}}{2Q_{*}}B_{(2)(3)}(r)\sin\theta.
\end{eqnarray}

To obtain the perturbations for the Schwarzschild solution we cannot just
set $Q_{*}$ equal to zero in Eqs. (69) and (70),
because we devided by $Q_{*}$ in several places during the development of
the formalism. However, Eqs. (37) and (38) are already decoupled for
the Schwarzschild black hole. This is such because in the Schwarzschild
spacetime the electromagnetic and gravitational fields are not coupled as
in the Reissner-Nordstr\"{o}m spacetime. Hence, one needs not decouple the
equations.

\section*{Acknowledgement}

It is a great pleasure for me to thank Amos Ori for many helpful discussions.

\end{document}